\documentclass{article}
\usepackage{spconf,amsmath,graphicx}
\usepackage{multirow}
\usepackage{comment}
\usepackage{booktabs}
\usepackage[textfont=it,tableposition=top]{caption}
\usepackage{color}
\usepackage{bm}
\pdfoutput=1
\newcommand{\PLH}{{\mkern-2mu\times\mkern-2mu}}

\title{From audio to semantics: \\Approaches to end-to-end spoken language understanding}

\name{Parisa Haghani\textsuperscript{*}, Arun Narayanan\sthanks{The first two authors have equal contribution, the rest of the list is in alphabetic order.}, Michiel Bacchiani, Galen Chuang}
\secondlinename{Neeraj Gaur, Pedro Moreno, Rohit Prabhavalkar, Zhongdi Qu, Austin Waters}
\address{Google Inc., USA}

\begin{document}
\ninept
\maketitle
\begin{abstract}
Conventional spoken language understanding systems consist of two main components: an automatic speech recognition module that converts audio to a transcript, and a natural language understanding module that transforms the resulting text (or top N hypotheses) into a set of domains, intents, and arguments. These modules are typically optimized independently. In this paper, we formulate audio to semantic understanding as a sequence-to-sequence problem \cite{sutskever2014sequence}. We propose and compare various encoder-decoder based approaches that optimize both modules jointly, in an end-to-end manner. Evaluations on a real-world task show that 1) having an intermediate text representation is crucial for the quality of the predicted semantics, especially the intent arguments and 2) jointly optimizing the full system improves overall accuracy of prediction. Compared to independently trained models, our best jointly trained model achieves similar domain and intent prediction $F1$ scores, but improves argument word error rate by 18\% relative.

\end{abstract}
\begin{keywords}
spoken language understanding, sequence-to-sequence, end-to-end training, multi-task learning, speech recognition
\end{keywords}
\section{Introduction}
\label{sec:intro}
Understanding semantics from a user input or a query is central to any human computer interface (HCI) that aims to interact naturally with users. Spoken dialogue systems that aim to solve this for specific tasks have been a focus of research for more than two decades \cite{tur2011spoken}. With the widespread adoption of smart devices like Google-Home \cite{li2017acoustic}, Amazon Alexa, Apple Siri and Microsoft Cortana, spoken language understanding (SLU) is moving to the forefront of HCI. 

\begin{figure}[th]
  \centering
  \includegraphics[width=\columnwidth]{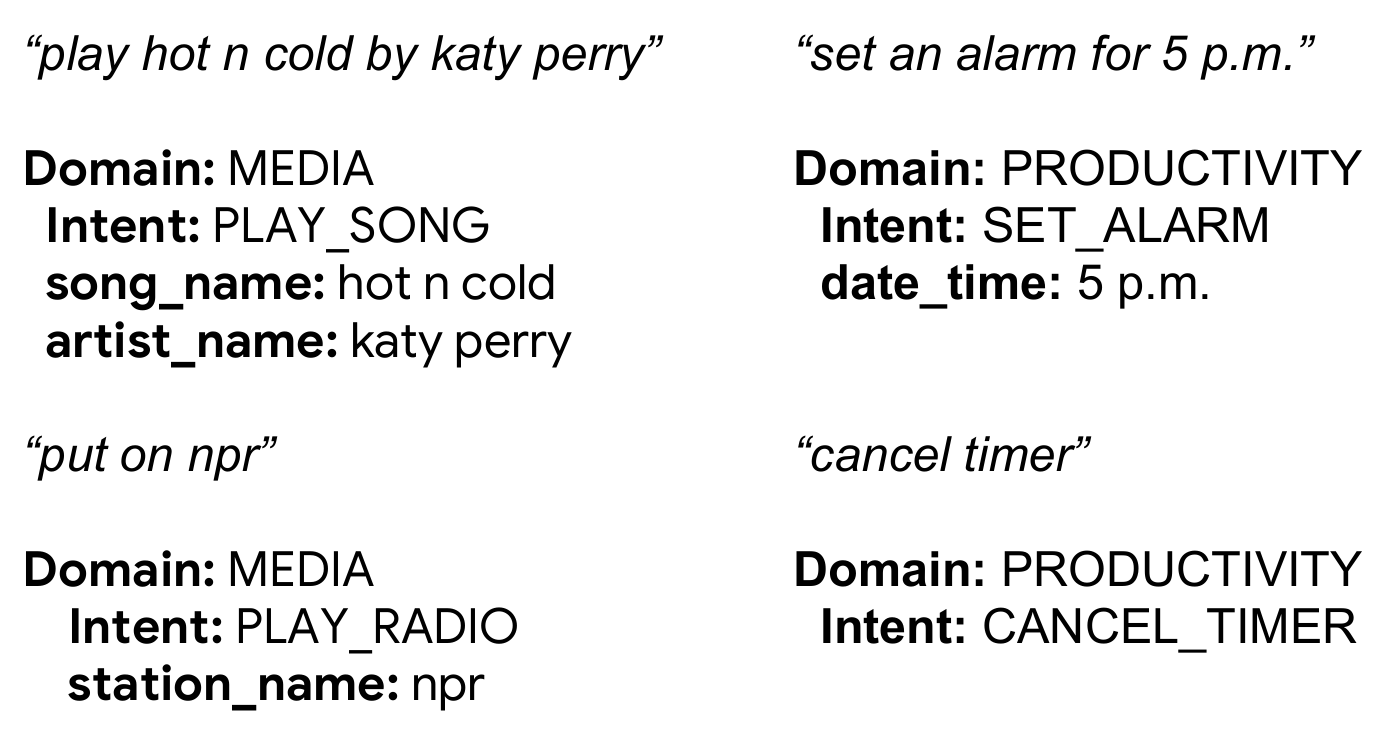}
  \caption{Example transcripts and their corresponding domain, intent, and arguments. Only arguments that have corresponding values in a transcript are shown. For example, song\_name and station\_name are both arguments in the MEDIA domain but only one has a corresponding value in each of the MEDIA examples.}
  \label{fig:nlu_examples}
\end{figure}

Typically, SLU involves multiple modules. An automatic speech recognition system (ASR) first transcribes the user query into a transcript. This is then fed to a module that does natural language understanding (NLU)\footnote{We use NLU to refer to the component that predicts domain, intent, and slots given the ASR transcript. SLU refers to the full system that predicts the aforementioned starting from audio.}. NLU itself involves domain classification, intent detection, and slot filling \cite{tur2011spoken}. In traditional NLU systems, first the high level domain of a transcript is identified. Subsequently, intent detection and slot filling are performed according to the predicted domain's semantic template. Intent detection identifies the finer-grained intent class a given transcript belongs to. Slot filling, or argument prediction,\footnote{We use slots and arguments interchangeably.} is the task of extracting semantic components, like the argument values corresponding to the domain. Figure~\ref{fig:nlu_examples} shows example transcripts and their corresponding domain, intent, and arguments. Recent work \cite{kim2017onenet, hakkani2016multi} has shown that jointly optimizing these three tasks improves the overall quality of the NLU component. For conciseness, we use the word \textit{semantics} to refer to all three of domain, intent and arguments.

Even though user interactions in an SLU system start as a voice query, most NLU systems assume that the transcript of the request is available or obtained independently. The NLU module is typically optimized independent of ASR. While accuracy of ASR systems have improved over the years \cite{Stolcke2017human, Saon2017swbd}, errors in recognition worsen NLU performance. This problem gets exacerbated on smart devices, where interactions tend to be more conversational. However, not all ASR errors are equally bad for NLU. For most applications, the semantics consist of an action with relevant arguments; a large part of the transcript of the ASR module has no impact on the end result as long as intent classification and predicted arguments are accurate. For example, for a user query, ``Set an alarm at two o'clock,'' intent, ``alarm,'' and its arguments, `two o'clock', are more important than filler words, like `an'. Joint optimization can focus on improving those aspects of transcription accuracy that are aligned with the end goal, whereas independent optimization fails at that objective. Furthermore, for some applications, there are intents that are more naturally predicted from audio compared to transcript. For example, when training an automated assistant, like Google Duplex \cite{duplex2018blog} or an airline travel assistant, it would be useful to identify acoustic events like background noise, music and other non-verbal cues as special intents, and tailor the assistant's response accordingly to improve the overall user experience. Hence, training various components of the SLU system jointly can be advantageous.

There have been some early attempts at using audio to perform NLU. Domain and intent are predicted directly from audio in \cite{serdyuk2018towards}, and this approach is shown to perform competitively, but worse than predicting from transcript. Alternatively, using multiple ASR hypothesis \cite{morbini2012reranking} or the word confusion network \cite{hakkani2006beyond} or the recognition lattice \cite{ladhak2016latticernn} have been proposed to account for ASR errors, but independent optimization of ASR and NLU can still lead to sub-optimal performance. In \cite{schumann2018}, an ASR correcting module is trained jointly with NLU component. To account for ASR errors, multiple ASR hypotheses are generated during training as additional input sequences, which are then error-corrected by the slot-filling model. While the slot-filling module is trained to account for the errors, the ASR module is still trained independent of NLU. Similar to the work in \cite{serdyuk2018towards}, an end-to-end system is proposed in \cite{ypchen2018} that does intent classification directly from speech, with an intermediate ASR task. But unlike the current work, it uses a connectionist temporal classification (CTC) \cite{graves2006connectionist} acoustic model, and only performs intent prediction. In the current work, we show that NLU, i.e., domain, intent, and argument prediction can be done jointly with ASR starting directly from audio and with a quality of performance that matches or surpasses an independently trained counterpart.

The systems presented in this study are motivated by the encoder-decoder based sequence-to-sequence (Seq2Seq) \cite{cho-al-emnlp14, bahdanau2014neural, sutskever2014sequence} approach that has shown to perform well for machine translation \cite{wu2016google} and speech recognition tasks \cite{chiu2017sota}. Encoder-decoder based approaches provide an attractive framework to implementing SLU systems, since the attention mechanism allows for jointly learning an alignment while predicting a target sequence that has a many-to-one relationship with its input \cite{chan2016listen}. Such techniques have already been used in NLU \cite{liu2016atten}, but using ASR transcripts, not audio, as input to the system.

In this work, we present and compare various end-to-end approaches to SLU for joinlty predicting semantics from audio. The presented techniques simplify the overall architecture of SLU systems. Using a large training set comparable to what is typically used for building large-vocabulary ASR systems, we show that not only can predicting semantics from audio be competitive, it can in some conditions outperform the conventional two-stage approach. To the best of our knowledge, this is the first study that shows all three of domain, intent, and arguments can be predicted from audio with competitive results.

The rest of the paper is organized as follows. Section~\ref{sec:archi} presents various models and architectures explored in this work. The experimental setup and results are described in Section~\ref{sec:experiments} and Section~\ref{sec:results}, respectively. We conclude in Section~\ref{sec:concl}.

\begin{figure}[th]
  \centering
  \includegraphics[width=\columnwidth]{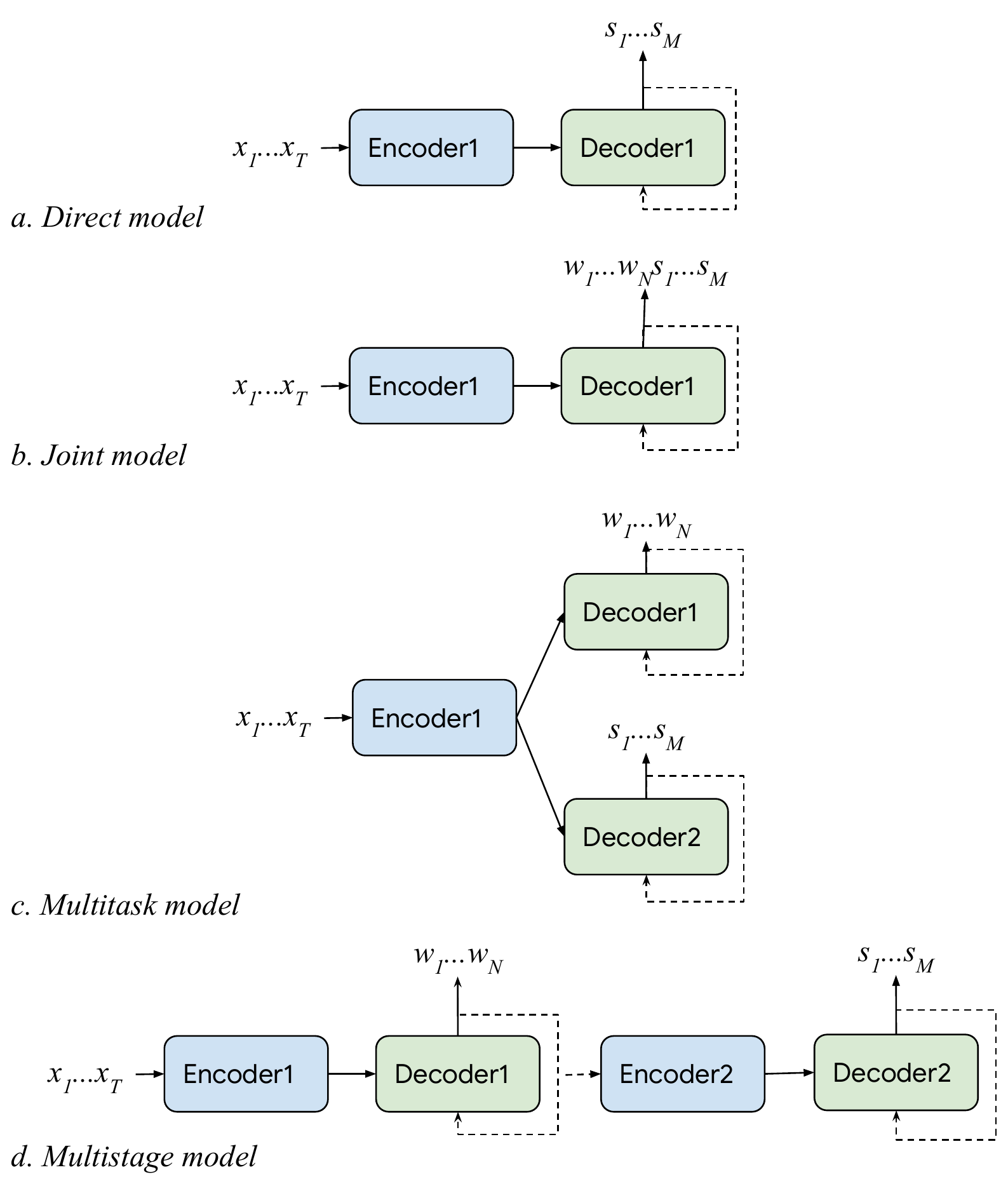}
  \caption{Different model architectures investigated in this paper. $\mathcal{X}$ stands for acoustic features, $\mathcal{W}$ for transcripts and $\mathcal{S}$ for semantics (domain, intent and arguments). Dotted lines represent both the conditioning of output label on its history and the attention module, which is treated as a part of the decoder.}
  \label{fig:architectures}
\end{figure}

\begin{table*}[t]
  \centering
  \begin{tabular}{ll}
   {\textbf{Transcript}} & {\textbf{Serialized Semantics}}\\
   \midrule
  ``can you set an alarm for 2 p.m.'' &
   \textless 
   DOMAIN\textgreater
   \textless
   PRODUCTIVITY\textgreater
   \textless
   INTENT\textgreater
   \textless
   SET\_ALARM\textgreater
   \textless
   DATETIME\textgreater 2 p.m.\\
   ``remind me to buy milk'' &
   \textless 
   DOMAIN\textgreater
   \textless
   PRODUCTIVITY\textgreater
   \textless
   INTENT\textgreater
   \textless
   ADD\_REMINDER
   \textgreater
   \textless
   SUBJECT\textgreater buy milk\\
   ``next song please''&
   \textless 
   DOMAIN\textgreater
   \textless
   MEDIA\_CONTROL\textgreater \\
   ``how old is barack obama''&
   \textless
   DOMAIN\textgreater
   \textless
   NONE\textgreater \\
   \bottomrule
  \end{tabular}
  \caption{Example transcripts and their corresponding serialized semantics.}
  \label{tab:home_examples}
\end{table*}

\section{System architecture}
\label{sec:archi}
Our work is based on the encoder-decoder framework augmented by attention. We start by reviewing this framework in Section~\ref{subsec:encdec}. 
There are multiple ways to model an end-to-end SLU system. One can either predict semantics directly from audio, ignoring the transcript, or have separate modules for predicting the transcript and semantics that are optimized jointly. These different approaches and the corresponding formulation are described in Sections~\ref{subsec:joint}~--~\ref{subsec:multistage}. Figure~\ref{fig:architectures} shows a schematic representation of these architectures.

\subsection{Notation}
\label{subsubsec:notation}
To review the general encoder-decoder framework, we denote the input and output sequences by \( \mathcal{A} = \{a_{1},\dots,a_{K}\} \) and \( \mathcal{B} = \{b_{1},\dots,b_{L}\} \), where $K$ and $L$ denote their lengths.
In this work, since we start from audio, the input sequence to the model are acoustic features (we describe the details of acoustic feature computation in Section~\ref{subsec:exp_models}), while the output sequence, depending on the model architecture, may be the transcript, the corresponding semantics, or both. While the semantics of an utterance is best represented as structured data, we use a simple deterministic scheme for serializing it by first including the domain and intent, followed by the argument labels and their values (see Table~\ref{tab:home_examples}). More details are described in Section~\ref{subsec:semantics_serialization}. For the rest of the paper, we denote the input acoustic features by \(\mathcal{X} = \{x_{1},\dots,x_{T}\}\), where $T$ stands for the total number of time frames. The transcript is represented as a sequence of graphemes. It is denoted as \(\mathcal{W} = \{w_{1},\dots,w_{N}\}\), where $N$ stands for the number of graphemes in the transcript. The semantics sequence is represented by \(\mathcal{S} = \{s_{1},\dots,s_{M}\}\), where $M$ stands for the number of tokens. The tokens come from a dictionary consisting of the domain, intent, argument labels, and graphemes to represent the argument values.

\subsection{Encoder-decoder framework}
\label{subsec:encdec}
Given the training pair \(\mathcal{(A,B)}\) and model parameters \(\theta\), a sequence-to-sequence model computes the conditional probability \(P(\mathcal{B|A; \theta})\). This can be done by estimating the terms of the probability using chain rule:

\begin{equation}\label{eq:e2e}
P(\mathcal{B|A; \theta}) = \prod\limits_{i=1}^{L}P(b_{i}|b_{1},\dots,b_{i-1},\mathcal{A;\theta})
\end{equation}
The parameters of the model are learned by maximizing the conditional probabilities for the training data:
\begin{equation}
\theta^\star = \underset{\theta}{\text{argmax}} \sum\limits_{(\mathcal{A,B})} \text{log} P(\mathcal{B|A; \theta})
\end{equation}

In the encoder-decoder framework \cite{cho-al-emnlp14, sutskever2014sequence}, the model is parameterized as a neural network, most commonly a recurrent neural network, consisting of two main parts: An \textit{encoder} that receives the input sequence and encodes it into a higher level representation, and a \textit{decoder} that generates the output from this representation after first being fed a special start-of-sequence symbol. Decoding terminates when the decoder emits the special end-of-sequence symbol. The modeling power of encoder-decoder framework has been improved by the addition of an \textit{attention} mechanism \cite{bahdanau2014neural}. This mechanism was introduced to overcome the bottleneck of having to encode the entire variable length input sequence in a single vector. At each output step, the decoder's last hidden state is used to generate an attention vector over the entire encoded input sequence, which is used to summarize and propagate the needed information from the encoder to the decoder at every output step. In this work, we use multi-headed attention \cite{vaswani2017attention} that allows the decoder to focus on multiple parts of the input when generating each output. The effectiveness of this type of attention for ASR was explored and verified in \cite{cc2018state_of_s2s}.

\subsection{Direct model}
\label{subsec:direct}
In the \textit{direct model} the semantics of an utterance are directly predicted from the audio. The model does not learn to fully transcribe the input audio; it learns to only transcribe parts of the transcript that appear as argument values. Conceptually, this is the simplest formulation for end-to-end semantics prediction. But it also makes the task challenging, since the model has to implicitly learn to ignore parts of the transcript that is not part of an argument and the corresponding audio, while also inferring the domain and intent in the process.

Following the notation introduced in Section \ref{subsec:encdec}, the model directly computes \(P(\mathcal{S|X; \theta})\), as in Equation~\ref{eq:e2e}. The encoder takes the acoustic features, \(\mathcal{X}\), as input and the decoder generates the semantic sequence, \(\mathcal{S}\).

\subsection{Joint model}
\label{subsec:joint}
This model still consists of an encoder and a decoder, similar to the direct model, but the decoder generates the transcript followed by domain, intent, and arguments. The output of this model is thus the concatenation of transcript and its corresponding semantics: \(\mathcal{[W:S]}\) where [:] denotes concatenation of the first and the second sequence.

This formulation conditions intent and argument prediction on the transcript:
\begin{equation}\label{eq:joint}
P(\mathcal{S,W|X; \theta}) = P(\mathcal{S|W,X; \theta})P(\mathcal{W|X; \theta})
\end{equation}
This model retains the simplicity of the direct model, while simultaneously making learning easier by introducing an intermediate transcript representation corresponding to the input audio.

\subsection{Multitask model}
\label{subsec:mtl}
Multitask learning \cite{caruana1997multitask} (MTL) is a widely used technique when learning related tasks, typically with limited data. Related tasks act as inductive bias, improving generalization of the main task by choosing parameters that are optimal for all tasks. Although predicting the text transcript is not necessary for domain, intent and argument prediction, it is a natural secondary task that can potentially offer a strong inductive bias while learning. In MTL, we factorize \(P(\mathcal{S,W|X; \theta})\) as:
\begin{equation}
P(\mathcal{S,W|X; \theta}) = P(\mathcal{S|X; \theta})P(\mathcal{W|X; \theta}).
\end{equation}
In the case of neural nets, multitask learning is typically done by sharing hidden representations between tasks \cite{luong2015multi}. In this work, we do this by sharing the encoder and having separate decoders for predicting transcripts and semantics. We then learn parameters that optimize both tasks:
\begin{equation}
\begin{split}
\theta^\star &= \\
&\underset{\theta}{\text{argmax}} 
\sum\limits_{(\mathcal{X,W,S})} 
\text{log} P(\mathcal{W|X; \theta_{\text{e}}, \theta_{\text{d}}^{W}})  +
\text{log} P(\mathcal{S|X; \theta_{\text{e}}, \theta_{\text{d}}^{S}}),
\end{split}
\end{equation}
where, \(\theta = (\theta_e, \theta_d^{W}, \theta_d^{S})\). \(\theta_e, \theta_d^{W}, \theta_d^{S}\) are the parameters of the shared encoder, the decoder that predicts the transcript, and the decoder that predicts semantics, respectively. The shared encoder learns representations that enable both transcript and semantics prediction. 

\subsection{Multistage model}
\label{subsec:multistage}
Multistage (MS) model, when trained under the maximum likelihood criterion, is most similar to the conventional approach of training the ASR and NLU components independently. In MS modeling, semantics are assumed to be conditionally independent of acoustics given the transcript:
\begin{equation}\label{eq:ms_loss}
P(\mathcal{S,W|X; \theta}) = P(\mathcal{S|W; \theta})P(W|X; \theta).
\end{equation}
Given this formulation, \(\theta\) can be learned as:
\begin{equation}
\begin{split}
\theta^\star &= \\
&\underset{\theta}{\text{argmax}} 
\sum\limits_{(\mathcal{X,W,S})} 
\text{log} P(\mathcal{W|X; \theta^{W}})  +
\text{log} P(\mathcal{S|X; \theta^{W}, \theta^{S}}),
\end{split}
\end{equation}
Here,  \(\theta^{W}, \theta^{S}\) are, respectively, the parameters of the first stage, which predicts the transcript, and the second stage, which predicts semantics. For each training example, we assume that the triplet \((\mathcal{X, W, S})\) is available. As a result, the two terms in Eq.~\ref{eq:ms_loss} can be independently optimized, thereby reducing the model to a conventional 2-stage SLU system. In practice, however, it is possible to weakly tie the two stages together during training by using the \emph{predicted} \(\mathcal{W}\) at each time-step and allowing the gradients to pass from the second stage to the first stage through that label index. In Sec.~\ref{sec:results}, we will present results using alternative strategies to pick \(\mathcal{W}\) from the first stage to propagate to the second stage, like the argmax of the softmax layer or sampling from the multinomial distribution induced by the softmax layer. By weakly tying the two stages, we allow the first stage to be optimized jointly with the second stage, based on the criterion that is relevant for both stages.

One of the advantages of the multistage approach is that the parameters for the 2 tasks are decoupled. Therefore, we can easily use different corpora to train each stage. Typically, the amount of data available to train a speech recognizer far exceeds the amount available to train an NLU system. In such cases, we can use the available ASR training data to tune the first stage and finally train the entire system using whatever data is available to train jointly. Furthermore, a stronger coupling between the 2 stages can be made when optimizing alternative loss criterion like the minimum Bayes risk (MBR) \cite{shannon2017optimizing}\cite{prabhavalkar2017minimum}. We'll leave these aspects of multistage modeling to future work, as the focus of current study is more to understand the feasibility of predicting directly from audio and training jointly. 

\section{Experimental setup}
\label{sec:experiments}
\subsection{Data}
Our training data consists of $24M$ anonymized English utterances transcribed by humans. Similarly, our test set consists of $16K$ hand-transcribed utterances. Both training and testing sets represent a slice of traffic from Google Home that we are interested in. The labeling for domain, intent, and arguments is generated from passing the ground-truth transcription through context free grammars (CFG). The CFGs are used to parse and transform ground-truth transcripts to domain, intent, and arguments. We only consider non-conversational (one-shot) queries in this work. In total, there are 5 domains: MEDIA, MEDIA\_CONTROL, PRODUCTIVITY, DELIGHT, and NONE. As the name suggests, any utterance that cannot be classified into the first four domains is labeled NONE. We consider $\sim$20 intents in this study, such as SET\_ALARM, SELF\_NOTE, etc., and two arguments: DATETIME and SUBJECT. The distribution of domains in the train and test sets is shown in Table~\ref{tab:google_home_domain_dist}.

\begin{table}[t]
  \caption{Distribution of domains considered in this study in the training and test data.}
  \label{tab:google_home_domain_dist}
  \centering
  \begin{tabular}{lll}
    \toprule
    {\textbf{domain}} & {\textbf{Train}} & {\textbf{Test}} \\
    \midrule
    MEDIA          &  $30\%$ & $20\%$ \\
    MEDIA\_CONTROL &  $8\%$ & $16\%$ \\
    PRODUCTIVITY   &  $7\%$ & $5\%$ \\
    DELIGHT        &  $2\%$ & $2\%$ \\
    NONE           &  $53\%$ & $56\%$ \\
    \bottomrule
  \end{tabular}
\end{table}

\subsection{Serializing/De-serializing Semantics}
\label{subsec:semantics_serialization}
We use a simple scheme for serializing semantics: The domain is specified first using a special tag `\textless{DOMAIN}\textgreater' followed by its name. If the domain is further divided into intents, we use the tag `\textless{INTENT}\textgreater' followed by the intent's name. Any optional arguments are specified similarly using the name of the argument and its corresponding value. Table~\ref{tab:home_examples} shows a few example transcripts and their corresponding serialized semantics.

At inference time, the predicted semantics sequence, \(\mathcal{S}\), is de-serialized in a similar fashion to extract the domain, intent, and argument label and values. This is done using a simple parser that tokenizes the sequence by the domain tag, intent tag and argument name and treats the sequence in between them as the corresponding values. This parser is agnostic to the order of these special tags, i.e., the domain tag can come ahead of the intent tag. In the case of the joint model where the output sequence is the concatenation of the transcript and semantics, the first observed  special tag or argument name marks the start of the semantic sequence.

The vocabulary that we use includes the domain and intent tags, domain, intent and argument names, (i.e.,
all symbols enclosed in ``\textless'' and ``\textgreater'' in Table~\ref{tab:home_examples}) as well as English graphemes for representing transcript and argument values. The graphemes in this study are limited to lowercase English alphabets and digits, punctuation and a few other special symbols such as underscore, brackets, start-of-sentence, and end-of-sentence. The total size of the vocabulary is $110$. Note that the special tags used for representing semantics are each a single ouput, e.g., ``\textless{DOMAIN}\textgreater'' is one output and not eight graphemes ``\textless , D, ..., N, \textgreater''.

\subsection{Models}
\label{subsec:exp_models}
All experiments use the same acoustic features: $80$-dimensional log-Mel filterbanks, computed with a $25$ msec window, shifted every $10$ msec. Similar to~\cite{sak2015lstm}, features from 3 contiguous frames are stacked, resulting in a $240$-dimensional vector. These stacked features are downsampled by a factor of 3 generating inputs at $30$ms frame rate that the encoder operates on.

\begin{table}[th]
    \caption{Model architectures used in the experiments. In each of the Enc/Dec columns, the first number indicates the number of layers and the second number shows the number of cells per layer. The cell type in all the models is Long Short Term Memory (LSTM). The last column shows the total number of parameters (in million).}
  \label{tab:arch}
  \centering
  \begin{tabular}{lllllr}
    \toprule
    \small{\textbf{Model}} & \small{\textbf{Enc.1}} & \small{\textbf{Dec.1}} & \small{\textbf{Enc.2}} & \small{\textbf{Dec.2}} & \small{\textbf{\#Params}} \\
    & (UniDi) & & (BiDi) & & \\
    \midrule
    Direct & \small{$5\PLH1400$} & \small{$2\PLH1024$} & - & - & \small{$97M$}\\
    Joint & \small{$5\PLH1400$} & \small{$2\PLH1024$} & - & - & \small{$97M$}\\
    Multitask & \small{$5\PLH1400$} & \small{$2\PLH512$} & - & \small{$2\PLH512$} & \small{$86M$}\\ 
    Multistage & \small{$5\PLH700$} & \small{$2\PLH512$} & \small{$5\PLH700$} & \small{$2\PLH512$} & \small{$84M$}\\
    \bottomrule
  \end{tabular}
\end{table}

\begin{table*}[t]
  \caption{Domain and intent $F1$ scores, and argument WER for the predicted semantics.} \label{tab:homeresult_recognized_semantics}
  \centering
  \begin{tabular}{llll}
    \toprule
    \textbf{Model} & \textbf{Domain F1} & \textbf{Intent F1} & \textbf{Arg WER} \\
    \midrule
    Baseline                   & $96.6$ & $95.1$ & $15.04$\\
    Direct                     & $96.2$ & $94.2$ & $18.22$\\
    Joint                      & \bm{$96.8$} & \bm{$95.7$} & $14.93$\\
    Multitask                  & \bm{$96.7$} & \bm{$95.8$} & $15.02$\\ 
    Multistage (ArgMax)         & $96.5$ & $95.4$ & $14.84$\\
    Multistage (SampledSoftmax) & $96.5$ & $95.2$ & \bm{$12.29$}\\
    \bottomrule
  \end{tabular}
\end{table*}

Table~\ref{tab:arch} summarizes the architecture of the various models used in our experiments. We maintain a similar number of parameters (within $~15\%$ difference) across models to allow for a fair comparison. All encoder and decoders use Long Short Term Memory (LSTM)~\cite{schmidhuber1997long} cells. The first encoder in all models is unidirectional, while the second encoder (in multistage models) uses bidirectional LSTMs~\cite{schuster1997bidirectional}.
Prior work ~\cite{hakkani2016multi} has shown that using bidirectional cells for encoding a transcript for the task of classifying its domain and intent achieves better performance compared to the unidirectional version. The first layer in all decoders is an embedding layer of size $128$. The second encoder in the multistage model, which takes the transcript as input, also uses an embedding layer of the same size. All decoders use $4$-headed additive attention~\cite{chiu2017state,bahdanau2014neural,vaswani2017attention}.

Our \textit{Baseline} is the multistage model in which the two stages that do ASR and NLU are trained independently, but using the same training data. We consider 2 variants of the multistage model that weakly couples the 2 stages. \textit{Multistage (ArgMax)} passes the argmax of the softmax layer of the first stage decoder, which predicts transcripts, to the second stage. \textit{Multistage (SampledSoftmax)}, on the other hand, passes on an unbiased sample from multinomial distribution represented by the output of the softmax layer \cite{bengio2015scheduled}. 

All neural networks are trained from scratch with the cross-entropy criterion in the TensorFlow framework~\cite{abadi2016tensorflow}. We use beam search during inference with a beam size of $8$. The models are trained using Tensor Processing Units \cite{jouppi2017datacenter} using the Adam optimizer \cite{kingma2014adam} and synchronous gradient descent. 

\subsection{Evaluation Metrics}
\label{subsec:home_metrics}
We use the typical ASR and NLU metrics for evaluation. For models that generate the transcript, we measure and report word error rate (WER). For semantics, we measure multi-class $F1$ scores \cite{CoNLL2003} for domain and intent. NLU systems that use in-out-begin (IOB) format for tagging arguments (see \cite{hakkani2016multi} for an example of IOB format) report $F1$ scores for argument tags (e.g., \cite{CoNLL2003} in the case of named-entities), but it is not clear how to measure this metric when the input transcript and the output arguments do not match, or when the input is audio. For example, if ground truth semantics contains ``\textless
DATETIME\textgreater
five p.m." but the hypothesized semantics is ``\textless
DATETIME\textgreater
high p.m.", it would be useful to have an error metric that captures the misrecognition of ``five" to ``high". For that reason, we choose to report WER for the arguments, instead of the $F1$ scores. In our computation, we count over triggers and misses towards $100\%$ WER. For example, if the ground truth semantics contains a DATETIME argument, but the recognized semantics does not, that instance has a $100\%$ WER for DATETIME. We compute per argument WER and report the weighted average where each argument's WER is weighted according to its number of occurrences.

\section{Results}
\label{sec:results}

Table~\ref{tab:homeresult_recognized_semantics} compares domain, intent, and argument prediction performance of the models presented in the previous section. As can be seen, all models perform relatively similarly when it comes to classifying the domains. The \textit{Joint} model works the best, with an $F1$ score of $96.8$\%. \textit{Direct} model, which has the lowest $F1$ score, is only worse by 0.6\% absolute. Performance on intent prediction is slightly worse, on average, compared to domain prediction. The \textit{Multitask} and \textit{Joint} models achieve the best $F1$ scores of $95.8$\% and $95.7$\%, respectively. Both these models use the encoded acoustic features as input to the decoder, and unlike the \textit{Direct} model, also predict the transcripts. This shows that having access to acoustic features and having an intermediate text representation are both important when predicting intent.

Comparing the \textit{Baseline} model with the multistage models that weakly couple the 2 stages, \textit{Multistage (ArgMax)} and \textit{Multistage (SampledSoftmax)}, we can see that they all work very similarly when it comes to domain and intent prediction, and are generally worse than \textit{Joint} and \textit{Multitask} models. This further shows the importance of complimenting transcripts with acoustic features when predicting intent.

The differences in argument WER is more pronounced among the different models. \textit{Direct} model performs the worst, getting a WER of $18.2$. This shows that including transcription loss while training end-to-end models can help improve argument prediction. Contrary to domain and intent $F1$ scores, \textit{Multistage (ArgMax)} and \textit{Multistage (SampledSoftmax)}, work better than the \textit{Joint} and \text{Multitask} models. Nevertheless, all jointly optimized models work better than the independently trained baseline. Notably, \textit{Multistage (SampledSoftmax)} model improves upon the baseline multistage model by $18$\% relative. 

Since the domain, intent and argument labeling for training and test data was obtained using CFG-parsers, we did a second experiment that used the predicted transcript from the various models, pipelined with the same CFG-parsers. The  CFG-parsers are used to derive domain, intent, and arguments from the predicted transcript. Results are shown in Table~\ref{tab:homeresult_transcript_and_parser}. The table also shows the overall WER obtained by the various models. Compared to the results in Table~\ref{tab:homeresult_recognized_semantics}, we can see that domain $F1$ scores are similar, but intent $F1$ scores are better. Interestingly, the argument WER significantly improved. For example, for the \textit{Baseline} model, WER improved from $15.0$ to $11.9$. While this is not entirely surprising, since this strategy of predicting semantics matches what is used for generating ground truth labels for training data, it is interesting to see that models that are optimized jointly still work better in terms of intent $F1$ scores and argument WER. For example, the \textit{Multitask} model gets an intent $F1$ score of $97.2$, which is better than the baseline by $1.3$ points. Similarly, \textit{Multistage (SampledSoftmax)} and \textit{Joint} models get an argument WER of $11.3$, which is $0.6$\% absolute better than the baseline. The results show that joint training can also help improve performance of the ASR component of the model when using the original CFG-parser for intent prediction.

\begin{table*}[th]
  \caption{Transcription WER, domain and intent $F1$ scores, and argument WER when NLU is performed on the model's top recognized transcript using the CFG-parser that was used for generating truth semantic labels during training.}
  \label{tab:homeresult_transcript_and_parser}
  \centering
  \begin{tabular}{lllll}
    \toprule
    \textbf{Model} & \textbf{WER} & \textbf{Domain F1} & \textbf{Intent F1} & \textbf{Arg WER} \\
    \midrule
    Baseline                   & $5.9$ & $96.4$ & $95.9$ & $11.89$\\
    Direct                     & -     & -      & -      & -\\
    Joint                      & \bm{$5.5$} & $96.5$ & $96.5$ & \bm{$11.28$}\\
    Multitask                  & $5.7$ & $95.4$ & \bm{$97.2$} & $11.71$\\ 
    Multistage (ArgMax)         & $5.9$ & $96.5$ & $96.3$ & $12.59$\\
    Multistage (SampledSoftmax) & $6.0$ & $96.5$ & $96.5$ & \bm{$11.26$}\\
    \bottomrule
  \end{tabular}
\end{table*}

\section{Concluding remarks}
\label{sec:concl}

In this work, we have proposed and evaluated multiple end-to-end approaches to SLU that optimize the ASR and NLU components of the system jointly. We show that joint optimization results in better performance not just when we do end-to-end domain, intent, and argument prediction, but also when using the transcripts generated by a jointly trained end-to-end model and a conventional CFG-parsers for NLU. Our results highlight several important aspects of joint optimization. We show that having an intermediate text representation is important when learning SLU systems end-to-end. As expected, our results also show that joint optimization helps the model focus on errors that matter more for SLU as evidenced by the lower argument WERs obtained by models that couple ASR and NLU. It was also observed that direct prediction of semantics from audio by ignoring the ground truth transcript, does not perform as well. 

There are several interesting avenues to improve performance going forward. As noted before, the amount of training data that is available to train ASR is usually several times larger than what is available to train NLU systems. It would be interesting to understand how a jointly optimized model can make use of ASR data to improve performance. For optimization, the current work uses the cross-entropy loss. Future work will consider more task specific losses, like MBR, that optimizes intent and argument prediction accuracy directly. It is also important to understand how to incorporate new grammars with limited training data into an end-to-end system. The CFG-parsing based approach that decouples itself from ASR can easily incorporate additional grammars. But end-to-end optimization relies on data to learn new grammars, making the introduction of new domains more challenging.

Framing spoken language understanding as a sequence to sequence problem that is optimized end-to-end significantly simplifies the overall complexity. It is also easy to scale such models to more complex tasks, e.g., tasks that involve multiple intents within a single user input, or tasks for which it is not easy to create a CFG-based parser. The ability to run inference without the need of additional resources like a lexicon, language models and parsers also make them ideal for deploying on devices with limited compute and memory footprint.

\section{Acknowledgements}
The authors would like to thank Edgar Gonz\`alez Pellicer, Alex Kouzemtchenko, Ben Swanson, Ashish Venugopal, Kai Zhao in help in obtaining labels used as truth for semantics, and Kanishka Rao for discussions around the joint model. We thank Khe Chai Sim and Amarnag Subramanya for helpful feedback on earlier drafts of the paper. 

\bibliographystyle{IEEEbib}


\end{document}